\documentclass[prd,superscriptaddress,nofootinbib,amsmath,amssymb,aps,11pt]{revtex4-2}

\usepackage{bm}
\usepackage{amsfonts}
\usepackage{latexsym}
\usepackage[latin1]{inputenc}
\usepackage{graphicx}
\usepackage{amsmath}
\usepackage{palatino}
\usepackage{mathpazo}
\usepackage[normalem]{ulem}

\usepackage{booktabs}
\usepackage{dcolumn}
\usepackage{natbib}
\usepackage[dvipsnames]{xcolor}

\begin{document}
	\title{Testing the limits of scalar-Gauss-Bonnet gravity through nonlinear evolutions of spin-induced scalarization}

    \author{Daniela D. Doneva}
	\email{daniela.doneva@uni-tuebingen.de}
	\affiliation{Theoretical Astrophysics, Eberhard Karls University of T\"ubingen, T\"ubingen 72076, Germany}

 	\author{Llibert Arest\'e Sal\'o}
	\email{l.arestesalo@qmul.ac.uk}
	\affiliation{School of Mathematical Sciences, Queen Mary University of London, Mile End Road, London, E1 4NS, United Kingdom}

  	\author{Katy Clough}
	\email{k.clough@qmul.ac.uk}
	\affiliation{School of Mathematical Sciences, Queen Mary University of London, Mile End Road, London, E1 4NS, United Kingdom}

 	\author{Pau Figueras}
	\email{p.figueras@qmul.ac.uk}
	\affiliation{School of Mathematical Sciences, Queen Mary University of London, Mile End Road, London, E1 4NS, United Kingdom}
	
	\author{Stoytcho S. Yazadjiev}
	\email{yazad@phys.uni-sofia.bg}
	\affiliation{Theoretical Astrophysics, Eberhard Karls University of T\"ubingen, T\"ubingen 72076, Germany}
	\affiliation{Department of Theoretical Physics, Faculty of Physics, Sofia University, Sofia 1164, Bulgaria}
	\affiliation{Institute of Mathematics and Informatics, Bulgarian Academy of Sciences, Acad. G. Bonchev St. 8, Sofia 1113, Bulgaria}

	\begin{abstract}
		Quadratic theories of gravity with second order equations of motion provide an interesting model for testing deviations from general relativity in the strong gravity regime. However, they can suffer from a loss of hyperbolicity, even for initial data that is in the weak coupling regime and free from any obvious pathology. This effect has been studied in a variety of cases including isolated black holes and binaries. Here we explore the loss of hyperbolicity in spin-induced scalarization of isolated Kerr black holes in a scalar-Gauss-Bonnet theory of gravity, employing the modified CCZ4 formulation that has recently been developed. We find that, as in previous studies, hyperbolicity is lost when the scalar field and its gradients become large, and identify the breakdown in our evolutions with the physical modes of the purely gravitational sector. We vary the gauge parameters and find the results to be independent of their value. This, along with our use of a different gauge formulation to previous works, supports the premise that the loss of hyperbolicity is dominated by the physical modes. Since scalar-Gauss-Bonnet theories can be viewed as effective field theories (EFTs), we also examine the strength of the coupling during the evolution. We find that at the moment when hyperbolicity is lost the system is already well within the regime where the EFT is no longer valid. This reinforces the idea that the theories should only be applied within their regime of validity, and not treated as complete theories in their own right.
	\end{abstract}
	
	\maketitle
	
	\section{Introduction}

	Black holes (BHs) are ideal objects for testing fundamental physics through electromagnetic and gravitational wave observations \cite{Berti:2015itd}. Firstly, because their compactness allows us to probe the highest curvature scales accessible in nature, and secondly, because they are remarkably simple objects, described only by their mass, charge and spin in the stationary case. 
    Even in the presence of additional fields, e.g., scalars, the Schwarzschild and the Kerr black holes are equilibrium solutions in a number of theories.
    This constitutes the basis of the no-scalar hair theorems \cite{heusler_1996}.   
	
	Theories of gravity that evade these theorems and are astrophysically relevant have attracted a lot of attention in recent years \cite{Herdeiro:2015waa}.
    Among the most prominent examples are effective field theories (EFTs) that can be viewed as a low-energy limit of quantum theories of gravity, containing an additional scalar degree of freedom with non-minimal couplings to curvature terms.  These theories may admit stationary BH solutions with non trivial hair, which can be used to impose observational constraints on the corresponding theories. For example, hairy BHs are the stationary end state in shift-symmetric or dilatonic Gauss-Bonnet gravity \cite{Kanti:1995vq,Torii:1996yi,Kleihaus:2011tg,Kleihaus:2015aje,Sotiriou:2013qea}, and in dynamical Chern-Simons gravity \cite{Yunes:2009hc,Delsate:2018ome,Yagi:2012ya}.  A further mechanism for generating BH hair that was recently identified is that of spontaneous scalarization \cite{Doneva:2017bvd,Silva:2017uqg,Antoniou:2017acq}, similar to neutron star spontaneous scalarization \cite{Damour:1993hw}  (see \cite{Doneva:2022ewd} for a review). This occurs in theories that coincide with GR in the weak field limit, but in which one can trigger a nonlinear development of the scalar field (i.e., a phase transition) when the compactness of the spacetime or the curvature reaches a certain threshold. Due to the fact that the sign of the Gauss-Bonnet term is spin-dependent close to the black hole horizon, in some cases, the trigger for scalarization can be a sufficiently rapid rotation of the black holes, rather than their mass \cite{Dima:2020yac,Doneva:2020nbb,Herdeiro:2020wei,Berti:2020kgk}. This latter effect has been dubbed ``spin-induced scalarization''.

	The construction of equilibrium solutions in such theories of gravity has advanced significantly in recent years, including the case of rapid rotation \cite{Cunha:2019dwb,Collodel:2019kkx,Herdeiro:2020wei,Berti:2020kgk}. In dynamical studies, the most widely used approach is the so-called decoupling limit approximation where the scalar field is evolved on top of the GR background (see e.g. \cite{Benkel:2016kcq,Benkel:2016rlz,Doneva:2021dqn,Kuan:2023trn,Richards:2023xsr} for the case of isolated objects and \cite{Okounkova:2017yby,Silva:2020omi,Elley:2022ept,Doneva:2022byd,Okounkova:2020rqw,Evstafyeva:2022rve} in the case of binaries). This method can give us important information about the scalar field dynamics and their astrophysical implications. Including the backreaction of the scalar field on the metric (and on the fluid in cases with matter) is a lower-order effect, and so is neglected in these works. It is, however, essential if we want to study in detail the impact of the scalar mode on the (tensor) gravitational waveforms. Thanks to recent developments in well-posed formulations \cite{Kovacs:2020pns,Kovacs:2020ywu}, such studies have recently begun to be carried out in parity invariant cases \cite{East:2021bqk,Corman:2022xqg,East:2022rqi,AresteSalo:2022hua,AresteSalo:2023mmd}. 
	
	One key reason for the slow progress in performing numerical evolutions of scalar-tensor EFTs is the question of well-posedness. Even when the field equations are of second order and the theory is free from ghosts, as in Gauss-Bonnet theories, it is not always clear that the theory can admit a well-posed formulation \cite{Ripley:2019hxt}. In particular, the question of well-posedness is dependent on the gauge and the initial conditions, although recent works have shown that in spherical symmetry the breakdown experienced is at least sometimes purely physical, and not related to the gauge choice \cite{R:2022hlf}.  In the weak coupling regime, where the EFT approach is justified, a Modified Generalized Harmonic Coordinate (mGHC) formulation has been identified for which strong hyperbolicity (and therefore well-posedness of the initial value problem) was proven in Lovelock and Horndeski theories of gravity \cite{Kovacs:2020ywu,Kovacs:2020pns}. The implementation of nonlinear evolutions within this mGHC formulation has since been performed in several works \cite{East:2020hgw,East:2021bqk,Corman:2022xqg}.
    The approach inspired a formulation within the alternative CCZ4 framework, using singularity-avoiding coordinates, which was proved to be well-posed in the weak coupling regime, and demonstrated to work in practice in \cite{AresteSalo:2022hua,AresteSalo:2023mmd}. This latter formulation is the one used in the present paper.

    However, even if for some choice of initial data the evolution is (locally) well-posed, there is no guarantee that the non-linearities of the theory will not drive the system into a strongly coupled regime, where local well-posedness need not hold. In fact, such a behavior should be expected (and is frequently observed in practice) since the propagation of the relevant fields is typically controlled by effective metrics that depend on the field values and their derivatives, just as in hydrodynamics, where shocks can form from smooth initial data.
	When moving away from the weak coupling regime one eventually encounters loss of hyperbolicity both for isolated black holes  \cite{Ripley:2019aqj,Ripley:2020vpk,Kuan:2021lol} and binaries \cite{East:2021bqk,East:2022rqi,AresteSalo:2022hua,AresteSalo:2023mmd}. The problems are not only due to the non-linearities of the theory -- loss of hyperbolicity is also observed when considering the perturbed field equations, although for a much narrower region in the parameter space \cite{Blazquez-Salcedo:2018jnn,Blazquez-Salcedo:2020rhf}. It is still not clear whether these problems can be cured by a proper choice of the gauge or if they are intrinsic to the theory, at least in the more general case beyond spherical symmetry \cite{R:2022hlf}. There are attempts to cure the loss of hyperbolicity by `fixing' the system of evolution equations \cite{Franchini:2022ukz,Cayuso:2023aht} or by modifying the theory, e.g., by adding an extra coupling of the scalar field to the Ricci scalar \cite{Thaalba:2023fmq,Liu:2022fxy}. 
	
	In the present paper, we explore the loss of hyperbolicity experienced following an exit from the weak coupling regime in a scalar-Gauss-Bonnet (sGB) theory of gravity that admits spontaneous scalarization. This problem has recently been studied in detail in spherical symmetry for the case of standard scalarization \cite{Ripley:2019hxt,R:2022hlf}. Here we study in detail the spin-induced case beyond spherical symmetry in 3+1D numerical relativity simulations using the modified CCZ4 gauge of \cite{AresteSalo:2022hua,AresteSalo:2023mmd}, expanding on previous work using the mGHC formulation in \cite{East:2021bqk}. Whilst the lower symmetry of this case compared to standard scalarisation makes the results more general, it means that, technically, it is more difficult to compute explicitly the propagation speeds of all the modes. However, we can still identify loss of hyperbolicity through two quantities related to the effective metric, and we study these, and the condition for weak coupling, in this work.

    The paper is organized as follows: Sec. II sets out the background for scalar-Gauss-Bonnet theories,  and Sec. III describes the modified CCZ4 formulation that is used. The criteria used to identify whether the theory is hyperbolic, and in the weakly coupling regime, are discussed in Sec. IV. Sec. V explains the coupling used, Sec. VI sets out numerical implementation details, and the results are presented in Sec. VII. We conclude in Sec VIII.

    \section{scalar-Gauss-Bonnet theory of gravity}

	We consider the scalar-Gauss-Bonnet theory having a general form of the action
	\begin{eqnarray}
		S=&&\frac{1}{16\pi}\int d^4x \sqrt{-g} 
		\Big[R - \frac{1}{2}\nabla_\mu \varphi \nabla^\mu \varphi - V(\varphi) 
		+ \frac{1}{4}\lambda\,f(\varphi){\cal R}^2_{GB} \Big] ,\label{eq:quadratic}
	\end{eqnarray}
	where $R$ is the Ricci scalar with respect to the spacetime metric $g_{\mu\nu}$,  ${\cal R}^2_{GB}$ is the Gauss-Bonnet invariant defined as ${\cal R}^2_{GB}=R^2 - 4 R_{\mu\nu} R^{\mu\nu} + R_{\mu\nu\alpha\beta}R^{\mu\nu\alpha\beta}$, $\varphi$ is the scalar field, and $V(\varphi)$ is the scalar field potential. The coupling between the scalar field and the Gauss-Bonnet invariant is controlled by the $ \lambda\,f(\varphi)$ coefficient, where $\lambda$ is a constant having dimensions of  $length^2$. For completeness, in this section, we include a scalar potential $V(\varphi)$, but in our simulations, we  set $V(\varphi)=0$ for simplicity.
	
	After varying the action with respect to the metric and the scalar field, one obtains the following field equations,
	\begin{eqnarray}\label{FE}
		&&R_{\mu\nu}- \frac{1}{2}R g_{\mu\nu} + \Gamma_{\mu\nu}= \frac{1}{2}\nabla_\mu\varphi\nabla_\nu\varphi -  \frac{1}{4}g_{\mu\nu} \nabla_\alpha\varphi \nabla^\alpha\varphi - \frac{1}{2} g_{\mu\nu}V(\varphi),\\
		&&\nabla_\alpha\nabla^\alpha\varphi= \frac{dV(\varphi)}{d\varphi} -  \frac{\lambda}{4} \frac{df(\varphi)}{d\varphi} {\cal R}^2_{GB},
	\end{eqnarray}
	where   $\Gamma_{\mu\nu}$ is defined as 	
	\begin{eqnarray}
		\Gamma_{\mu\nu}&=& -\frac{1}{2}R\Omega_{\mu\nu} - \Omega_{\alpha}^{~\alpha}\left(R_{\mu\nu} - \frac{1}{2}R g_{\mu\nu}\right) 
                + 2\,R_{\alpha(\mu}\Omega^{~\alpha}_{\nu)} - g_{\mu\nu} R^{\alpha\beta}\Omega_{\alpha\beta} 
		+ \,  R^{\beta}_{\;\mu\alpha\nu}\Omega^{~\alpha}_{\beta}\,
	\end{eqnarray}  
	with 	
	\begin{eqnarray}
		\Omega_{\mu\nu}= \lambda\,\nabla_{\mu}\nabla_{\nu}f(\varphi) .
	\end{eqnarray}

    In order to study spin-induced scalarisation, we need to choose a coupling function $f(\varphi)$ that is quadratic to first order. The exact form we use is given in Sec. \ref{sec-coupling}.

    \section{Modified CCZ4 formulation}
	
    In our numerical evolutions we employ the modified CCZ4 formulation of \cite{AresteSalo:2022hua,AresteSalo:2023mmd}, which was proven to lead to a strongly hyperbolic system of equations in the weakly coupled regime for the four-derivative scalar-tensor theory of gravity ($4\partial$ST) (which includes the scalar-Gauss-Bonnet theory as a subset). Here we will summarise the key points of the method - see \cite{AresteSalo:2022hua,AresteSalo:2023mmd} for further details.

    In this formulation the equations are rendered well-posed by adding a set of terms that vanish when the constraints hold ($Z_{\mu}=0$) to the equations of motion of the metric part in \eqref{FE}, as in \cite{Alic:2011gg}. Specifically, we perform the following replacement,
    \begin{eqnarray}\label{mod_FE}
        R^{\mu\nu}-\tfrac{1}{2}R g^{\mu\nu}\to R^{\mu\nu}-\tfrac{1}{2}R g^{\mu\nu}+2\big(\delta_{\alpha}^{(\mu}\hat{g}^{\nu)\beta}-\tfrac{1}{2}\delta_{\alpha}^{\beta}\hat{g}^{\mu\nu}\big)\nabla_{\beta}Z^{\alpha}-\kappa_1\big[2n^{(\mu}Z^{\nu)}+\kappa_2n^{\alpha}Z_{\alpha}g^{\mu\nu} \big]\,,
    \end{eqnarray}
    where $\hat{g}^{\mu\nu}$ and $\tilde{g}^{\mu\nu}$ are two auxiliary Lorentzian metrics that ensure that gauge modes and gauge condition violating modes propagate at distinct speeds from physical modes, as in \cite{Kovacs:2020pns, Kovacs:2020ywu} \footnote{Note that $\tilde{g}^{\mu\nu}$ is hidden in the definition of the constraints $Z^{\mu}$ (see \cite{AresteSalo:2022hua,AresteSalo:2023mmd} for further details).}. They can be defined as
    \begin{eqnarray}\label{eq:guage_choice}
        \tilde{g}^{\mu\nu}=g^{\mu\nu}-a(x)n^{\mu}n^{\nu}\,\qquad \hat{g}^{\mu\nu}=g^{\mu\nu}-b(x)n^{\mu}n^{\nu}\,,
    \end{eqnarray}
    where $a(x)$ and $b(x)$ are arbitrary functions such that $0<a(x)<b(x)$ and $n^{\mu}=\tfrac{1}{\alpha}(\delta_t^{\mu}-\beta^i\delta_i^{\mu})$ is the unit timelike vector normal to the $t\equiv x^0 = $const. hypersurfaces with $\alpha$ and $\beta^i$ being the lapse function and shift vector of the $3+1$ decomposition of the spacetime metric, namely
    \begin{eqnarray}
        ds^2=-\alpha^2dt^2+\gamma_{ij}(dx^i+\beta^idt)(dx^j+\beta^jdt)\,.
    \end{eqnarray}
    Note that $a(x)=b(x)=0$ leads to the usual CCZ4 formulation \cite{Alic:2011gg}, including the damping terms in \eqref{mod_FE}, whose coefficients $\kappa_1>0$ and $\kappa_2>-\tfrac{2}{2+b(x)}$ guarantee that constraint violating modes are exponentially suppressed \cite{AresteSalo:2022hua,AresteSalo:2023mmd}. In most of our simulations, we set $a(x)=0.2$ and $b(x)=0.4$ as in \cite{East:2021bqk}, but have also explored the sensitivity of our results to changes in these values, as detailed below.  

    In \cite{AresteSalo:2022hua,AresteSalo:2023mmd} the evolution equations for the $3+1$ formalism are derived and specified, and we do not reproduce them here. The versions of the $1+log$ slicing and Gamma-driver evolution equations that result in the modified puncture gauge are
    \begin{eqnarray}
        \partial_t\alpha=\beta^i\partial_i\alpha-\tfrac{2\alpha}{1+a(x)}(K-2\Theta)\,,\\
        \partial_t\beta^i=\beta^j\partial_j\beta^i+\tfrac{3}{4}\tfrac{\hat{\Gamma}^i}{1+a(x)}-\tfrac{a(x)\alpha\partial_i\alpha}{1+a(x)}\,,
    \end{eqnarray}
    where $\Theta=Z^0$, $K$ is the trace of the extrinsic curvature of the induced metric $\gamma_{ij}$, and $\tilde{\Gamma}^i=\tilde{\gamma}^{kl}\tilde{\Gamma}^i_{kl}$, with $\tilde{\Gamma}^i_{kl}$ being the Christoffel symbols associated to the conformal spatial metric $\tilde{\gamma}_{ij}\equiv\chi\gamma_{ij}$, where $\chi=\det(\gamma_{ij})^{-1/3}$.

    \section{Weak coupling regime and well-posedness}\label{sec:wfc}

    The formulation of \cite{AresteSalo:2022hua,AresteSalo:2023mmd} has been proven to be strongly hyperbolic for sGB (and thus well-posed) in the weakly coupled regime, where the contributions of the Gauss-Bonnet term to the field equations, measured by the coupling $\sqrt{\lambda f'(\varphi)}$, are smaller than the two-derivative Einstein-scalar field terms. This yields the following {\emph Weak Coupling Condition} (WCC),
    \begin{eqnarray}\label{wfc}
        \sqrt{|\lambda\,f'(\varphi)|}/L\ll 1\,,
    \end{eqnarray}
    where $L^{-1}=\max\{|R_{ij}|^{1/2}, |\nabla_{\mu}\varphi|, |\nabla_{\mu}\nabla_{\nu}\varphi|^{1/2},|{\cal R}^2_{GB}|^{1/4}\}$ is the inverse of the shortest physical length scale characterizing the system, i.e., the maximum curvature scale.

    In this work, we explore the validity of the EFT by monitoring the condition \eqref{wfc} during the evolution. This condition is, in a sense, more important than hyperbolicity, since once breached one can no longer trust the EFT. Nevertheless, it is interesting to explore the interplay between the two conditions and how closely they coincide in practice for generic classes of initial conditions.

    It was shown in \cite{AresteSalo:2022hua,AresteSalo:2023mmd} that the contribution from the Gauss-Bonnet sector to the principal part of the evolution equations only affects the physical modes (and not the gauge modes). These can be separated into purely gravitational modes and mixed scalar-gravitational ones \cite{Reall:2021voz}. Therefore, strong hyperbolicity (and thus, well-posedness) fails when the eigenvalues corresponding to those modes become imaginary.

    In \cite{AresteSalo:2022hua,AresteSalo:2023mmd} all the physical eigenvalues were computed perturbatively. However, the eigenvalues from the purely gravitational sector can be derived exactly in the full theory, given that they lie on the null cone of the effective metric \cite{Reall:2021voz},
    \begin{eqnarray}
        g_{\text{eff}}^{\mu\nu}=g^{\mu\nu}-\Omega^{\mu\nu}\,.
    \end{eqnarray}
    Hence, one can find that the determinant of the effective metric (normalized to its value in pure GR) is given by
    \begin{eqnarray}\label{disc}
        \frac{\det(g^{\mu\nu}_{\text{eff}})}{\det(g^{\mu\nu})}=\left(\tfrac{1}{1+\Omega^{\perp\perp}}\right)^2\det\left\{\tfrac{1}{\chi}\big[(\gamma^{ij} - \Omega^{ij})(1 + \Omega^{\perp\perp}) -\tfrac{2}{\alpha}\Omega^{\perp(i} \beta^{j)} - \Omega^{\perp\perp} \tfrac{\beta^i \beta^j}{\alpha^2}+ \Omega^{\perp i} \Omega^{\perp j}\big]\right\}\,,
    \end{eqnarray}
    where $\Omega^{ij}=\gamma^i_{\mu}\gamma^j_{\nu}\Omega^{\mu\nu}$, $\Omega^{\perp i}=-n_{\mu}\gamma^i_{\nu}\Omega^{\mu\nu}$ and $\Omega^{\perp\perp}=n_{\mu}n_{\nu}\Omega^{\mu\nu}$. When the value of the ratio in \eqref{disc} becomes negative, this tells us that strong hyperbolicity no longer holds. The eigenvalues of these modes of the system have become imaginary, and when this occurs outside of the apparent horizon the evolution cannot continue. It is also possible that in the strongly coupled regime, strong hyperbolicity could be violated in the mixed scalar-gravitational sector; in this case the diagnosis of the problem is less easy to formulate
    \footnote{In this paper we only identify when loss of hyperbolicity is observed in the purely gravitational modes. Since the loss of hyperbolicity in these modes appears to coincide with the breakdown of the simulation, we do not investigate further the mixed scalar-gravitational ones. Whilst the latter are the ``fastest'' modes \cite{Reall:2021voz}, it is not necessarily the case that hyperbolicity loss should occur first in their sector. Further work is needed to understand at what level they contribute to the loss of hyperbolicity we observe.}.

    When hyperbolicity is lost, as discussed in \cite{Bernard:2019fjb} (see also \cite{Figueras:2020dzx}), the equations change character from hyperbolic to parabolic or elliptic. This behavior can be described by analogy with two model equations, namely the Tricomi equation,
    \begin{eqnarray}
        \partial_y^2u(x,y)+y\partial_x^2u(x,y)=0\,,
    \end{eqnarray}
    where the characteristic speeds go to zero at $y=0$ and the equations become parabolic, or the Keldysh equation,
    \begin{eqnarray}
        \partial_y^2u(x,y)+\tfrac{1}{y}\partial_x^2u(x,y)=0\,,
    \end{eqnarray}
    where the characteristic speeds diverge at the transition line $y=0$. Finding out if loss of hyperbolicity happens due to a Tricomi or Keldysh-type transition is of interest since the cure is different for each type. For instance, one can choose a different gauge for a Keldysh-type transition or add derivative self-interactions for a Tricomi-type transition, see the discussion in \cite{Barausse:2022rvg} for more details. 
	
	\section{Spin-induced scalarization and the coupling function}
    \label{sec-coupling}
 
	The specific choice of sGB theory is controlled by the coupling function $f(\varphi)$ and the potential. In what follows we will assume $V(\varphi)=0$.
	
	Loss of hyperbolicity has been considered in several classes of sGB gravity, including the shift-symmetric case with $f(\varphi)=\varphi$ \cite{Ripley:2019aqj,R:2022hlf} as well as the quadratic case, $f(\varphi)=\varphi^2$, which allows for spontaneous scalarization in black hole spacetimes \cite{Ripley:2020vpk,R:2022hlf}. The first results for spin-induced scalarization were reported in \cite{East:2021bqk}, where quadratic-quartic ($f(\varphi)=\tfrac{\lambda}{2}\varphi^2+\tfrac{\sigma}{4}\varphi^4$) and exponential ($f(\varphi)=\tfrac{\lambda_e}{6}(1-e^{-3\varphi^2})$) couplings were considered. In the present paper, we further explore the loss of hyperbolicity for spin-induced scalarization scenarios, investigating the type of hyperbolicity loss, and the role played by the gauge choice. 
 
    Spin induced scalarization happens only for sufficiently rapidly rotating BHs and a coupling that is quadratic at leading order, with $\lambda<0$, see eq. \eqref{eq:quadratic}.
    If a subclass of sGB gravity admits spontaneous scalarization, then Kerr is always a solution of the sGB field equations. The scalar field development is sourced by the Gauss-Bonnet invariant -- considering the scalar field perturbations in \eqref{FE} on top of the GR background, one obtains
	\begin{equation}
		\left( \nabla_\alpha\nabla^\alpha - m^2_{\rm eff} \right)\delta \varphi= 0\,,
	\end{equation}
    where the scalar field effective mass is defined as 
    \begin{equation}
        m^2_{\rm eff} = - \frac{\lambda}{4} \left.\frac{d^2 f(\varphi)}{d\varphi^2 }\right|_{\varphi=0} {\cal R}^2_{GB}\,.
    \end{equation}
    When $m^2_{\rm eff}<0$ there is a tachyonic instability, and the scalar field grows exponentially until it is quenched by nonlinear effects. For static black holes, we always have ${\cal R}^2_{GB}>0$ and, thus, scalarization is only possible when $\lambda\,d^2 f(\varphi)/d\varphi^2|_{\varphi=0}>0$. On the other hand, for rapidly rotating Kerr black holes with $a/M>0.5$,  ${\cal R}^2_{GB}$ becomes negative close to the poles around the black hole horizon. In such cases, a coupling function with $\lambda\,d^2 f(\varphi)/d\varphi^2|_{\varphi=0}<0$ can lead to negative $ m^2_{\rm eff}$ in some regions. For a given black hole mass and large enough $\lambda$ (the exact threshold depends on the value of $a$) the scalar field perturbations will become unstable, leading to the development of scalar hair around the black hole \cite{Dima:2020yac} \footnote{Note that in the presence of matter, e.g., for neutrons stars, we can have scalarization for both signs of $\lambda\,d^2 f(\varphi)/d\varphi^2$ even in the static case.}. 
    
	The simplest coupling function that can satisfy the conditions for spin-induced scalarization is  $f(\varphi)=\varphi^2$ with $\lambda<0$. However, this leads to unbounded instabilities, which are undesirable (from a numerical point of view) -- they cannot be followed to a stable end state\footnote{For other types of instabilities potentially present for scalarized black holes see \cite{Minamitsuji:2023uyb}.}. We therefore choose an exponential form of the coupling, namely
	\begin{equation}
		f(\varphi)=  \frac{1}{2 \beta} \left[1-\exp(-\beta \varphi^2)\right], \label{eq:coupling_function}
	\end{equation} 
	where $\beta$ is a constant. In the vicinity of $\varphi=0$ this coupling has the same leading order expansion as the pure quadratic coupling. Thus, the phenomenology of spontaneous scalarization in both cases is very similar. However, the exponential term introduces higher-order corrections that lead to a saturation of the instability above a certain amplitude of the scalar field. In particular, with this choice, the value of the dimensionless coupling  $\lambda/M^2$ controls whether, for a fixed initial $a_0/M$, a given Kerr black hole can scalarize and, if so, the timescale of the growth of the scalar cloud. The constant $\beta$ in the coupling function \eqref{eq:coupling_function} controls the scalar field amplitude in the final equilibrium state; namely, larger $\beta$ leads to a weaker scalar field in the final state. In \cite{East:2021bqk} the exponential coupling studied corresponded to a fixed value of $\beta=6$, whereas we allow this parameter to vary.

    \section{Numerical implementation and methods}
    \label{sec-numericsetup}
    In order to study the loss of hyperbolicity and its relation to violation of the weak coupling condition we have performed $3+1$ nonlinear evolutions of spin induced scalarisation for isolated rotating black holes. We use the \texttt{GRChombo} code \cite{Andrade:2021rbd,Radia:2021smk} and more specifically its modification to include sGB gravity with singularity avoiding coordinates \cite{AresteSalo:2022hua,AresteSalo:2023mmd}, in contrast to other recent studies that make use of mGHC \cite{East:2021bqk}. Using different gauge formulations helps us to explore the possibility that these are physical breakdowns in the theory and not related to gauge issues.

    The size of the computational domain is $L=256M$ along each of the coordinate directions and we use 6 refinement levels (so 7 levels in total), with a refinement ratio of $2:1$. The rest of the parameters are fixed to $\kappa_1=2.0/M$, $\kappa_2=-0.1$, while the Kreiss-Oliger numerical dissipation coefficient is set to  $\sigma=2.0$ (see \cite{Radia:2021smk} for the precise definition of these parameters).
    
    \begin{figure}[]
        \includegraphics[width=0.45\textwidth]{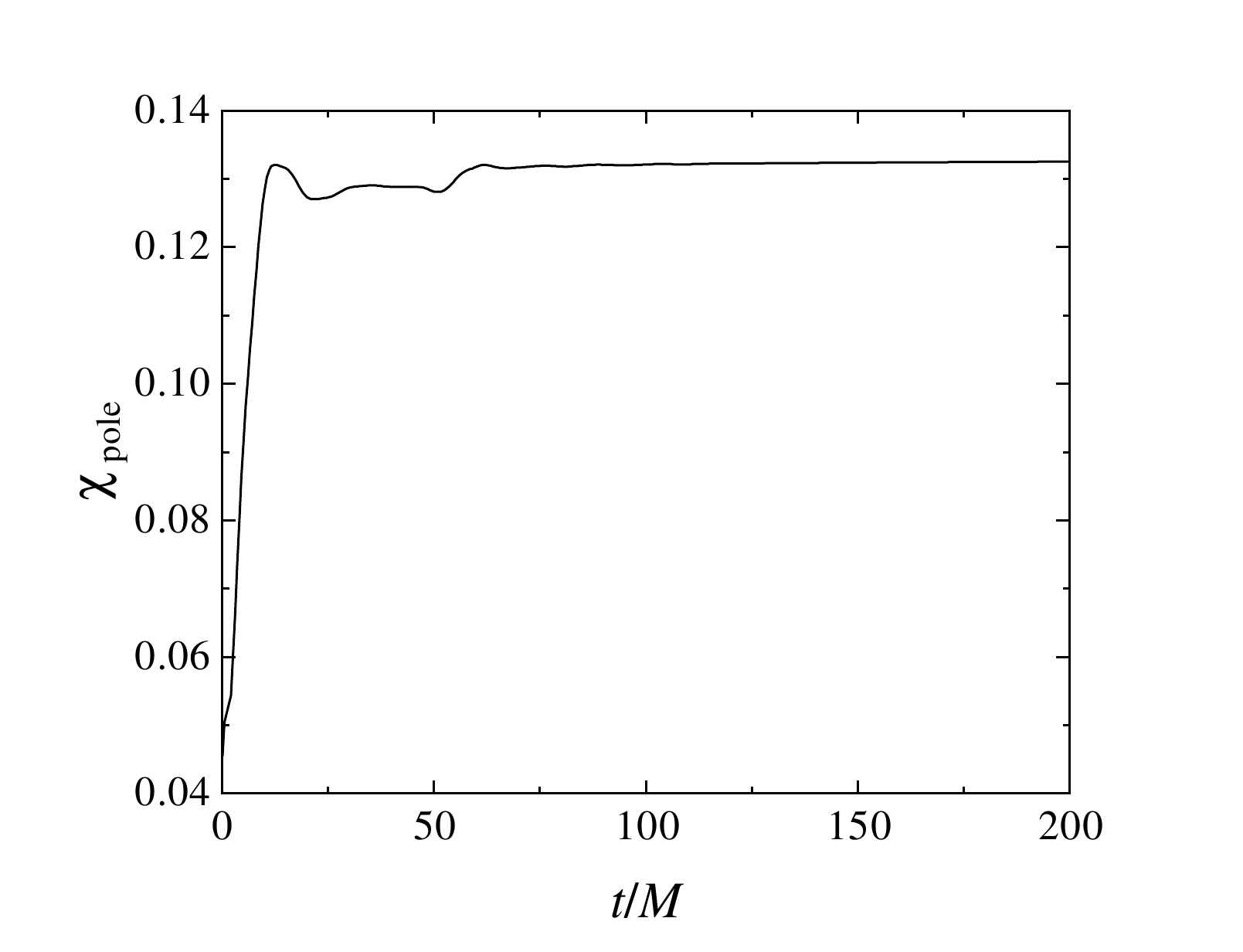}
        \includegraphics[width=0.45\textwidth]{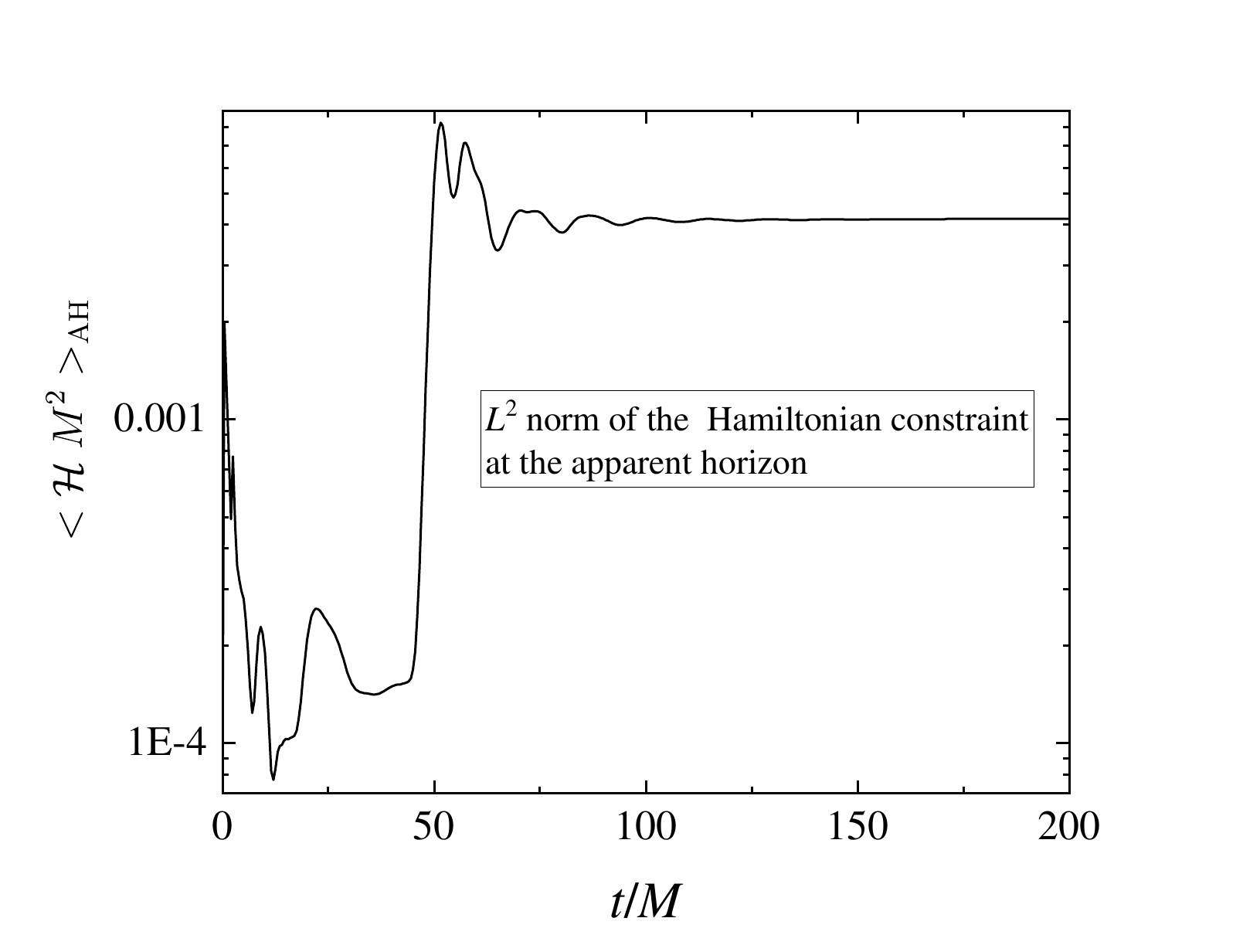}
        \caption{Time evolution of the conformal factor $\chi$ at the pole of the apparent horizon \textit{(left panel)} and the $L^2$ norm of the Hamiltonian constraint at the apparent horizon \textit{(right panel)}. The simulations are performed for $a_0/M=0.8$, $\lambda/M^2=-50$, and $\beta=500$. We see that the gauge evolution has roughly settled by $t=30M$, as has the initial constraint violation on the horizon. At this point, the scalar pulse hits the BH, and the scalar field starts to grow strongly just before $t=50M$. This is associated with further evolution of the metric and some growth in the constraints, but this stabilises when the field saturates the instability.}
        \label{fig:Evol_chi_Ham}
    \end{figure}
    
    Our initial conditions are isolated Kerr black holes in quasi-isotropic coordinates (see \cite{Liu:2009al} for details) with initial angular momentum parameters $a_0/M=0.6$ and $a_0/M=0.8$ respectively. On top of the GR background, we add a scalar field perturbation in the form of a small Gaussian pulse located at a distance of $30M$ from the center of the black hole with an amplitude of $10^{-5}$ and zero initial momentum. Due to the use of the puncture gauge, our initial data is not a stationary solution of the 3+1 evolution equations and the variables need a characteristic time of $\sim 10-20M$ to settle to a nearly stationary state. This is demonstrated in the left panel of Fig. \ref{fig:Evol_chi_Ham} where the evolution of the conformal factor $\chi$ is plotted. This figure shows that $\chi$ grows on a timescale of $\sim 10-20M$ and reaches a roughly constant state before the scalarization begins.
    We note that the addition of the scalar pulse violates the constraints by a small amount, and we rely on the constraint damping terms in the equations of motion to get rid of this error during the evolution.  However, for our choice of scalar amplitude the initial constraint violations are small compared to the truncation errors introduced by the discretization of the equations. 
    By the time the scalar pulse ``hits'' the black hole, the initial period of gauge adjustment is over and the constraints are well satisfied, as we can see in Fig. \ref{fig:Evol_chi_Ham}. Here the Hamiltonian constraint is
    \begin{eqnarray}
        \mathcal{H}= H-\tfrac{1}{2}(K_{\phi}^2+(\partial_i\phi)^2)-\Omega H+2\,H_{kl}\Omega^{kl}\,,
    \end{eqnarray}
    where $H_{ij}=R_{ij}+KK_{ij}-K_{ik}K_j^{~k}$, $\Omega_{ij}=\gamma_i^{\mu}\gamma_j^{\nu}\Omega_{\mu\nu}$, $H=\gamma^{ij}H_{ij}$ and $\Omega=\gamma^{ij}\Omega_{ij}$. 
	
	In our simulations, $|\lambda/M^2|$ is chosen large enough such that the corresponding Kerr black hole undergoes spin-induced scalarization.
    The small scalar field perturbation grows exponentially until it settles into an equilibrium distribution. In an astrophysical setup, this exponential growth would happen during stellar core collapse to a black hole. In that case, the curvature quickly grows as the stellar core compactifies and this would trigger the scalar field development \cite{Kuan:2021lol}. Thus, whilst the setup we consider is somewhat artificial, it is a reasonable proxy for the astrophysical processes that could produce scalarized black holes.
    As we will see below, the strongest violation of the hyperbolicity and weak coupling condition happens during the exponential growth of the scalar field in the intermediate stages of evolution, so this is the phase that should constrain theories most strongly.
	
	As already observed in $1+1$ nonlinear simulations in the spherically symmetric case, the non-hyperbolic region first develops inside the black hole horizon. As the evolution proceeds it can emerge above the horizon leading to an ill-posed initial value problem \cite{Corelli:2022phw}. Even though a non-hyperbolic region would not, in principle, be a problem from a physical point of view if it is hidden inside the apparent horizon, from a numerical point of view it still leads to problems continuing the evolution in a horizon-penetrating puncture gauge (in the mGHC approach one excises the region inside the black hole horizon from the evolution domain \cite{Ripley:2019hxt,Ripley:2020vpk}). We fix this by changing the equations of motion inside the apparent horizon, smoothly switching off the coupling constant as in \cite{Figueras:2020dzx,Figueras:2021abd, AresteSalo:2022hua,AresteSalo:2023mmd}. This means that our interior is effectively Kerr and thus hyperbolic, while outside the horizon the full sGB system of equations is evolved. This approach is justified as long as we turn off the Gauss-Bonnet term fully inside the horizon and the details of how we switch off this term do not influence the physics of the system on and outside the horizon. 
	To achieve this, the coupling function that we have employed in practice has the form
	\begin{equation} \label{eq:excision}
		f(\varphi) =  f_{\rm orig}(\varphi) /\left(1 + e^{-\beta_{\rm ex} (r - r_{\rm ex})}\right)
	\end{equation}  
    where $f_{\rm orig}$ is the original coupling defined in \eqref{eq:coupling_function}. In our simulations, we have set  $\beta_{\rm ex}=400$ while $r_{\rm ex}$ is a parameter smaller than the calculated black hole apparent horizon radius. For relatively weak scalar fields, the value of $r_{\rm ex}$ would depend mostly on the initial $a_0/M$.  We have checked that the scalar field evolution outside the apparent horizon remains unchanged (if hyperbolic) with the decrease of  $r_{\rm ex}$. Since, as discussed above, unstable regions often develop within the horizon and gradually extend beyond it, then in order to determine the limiting parameters for loss of hyperbolicity in the exterior of the BH we have to choose the maximum possible $r_{\rm ex}$ inside the apparent horizon. Thus, for initial $a_0/M=0.6$ we have worked with $r_{\rm ex}=0.75M$ while for $a_0/M=0.8$, the excision radius $r_{\rm ex}=0.54M$.
	
    A final comment on $r_{\rm ex}$ concerns the fact that the apparent horizon radius for the final rotating black holes states may not be exactly spherical after the gauge evolution. 
    Thus, using a spherical excision radius might introduce an error in the measured values. Our aim in the paper is to make an order of magnitude estimate of the threshold for hyperbolicity loss and the validity of the weak coupling condition, and so we consider our approach to be appropriate for this goal.

    \section{Results}
    
    \subsection{Hyperbolicity loss threshold and weak coupling condition}
 
	In order to determine the hyperbolicity loss threshold, for each value of $a_0/M$ we have performed a series of simulations where we vary the parameters $\lambda/M^2$ and $\beta$ in turn.
    The goal is to determine the threshold of $\beta$ for each combination of $a_0/M$ and $\lambda/M^2$ where loss of hyperbolicity is observed. The results are summarized in Table \ref{tab:threshold}. We see that $\beta_{\rm threshold}$ is typically of the order of $10^2-10^3$ for values of $|\lambda/M^2|$ close to the minimum value that ensures scalarization. The value of $\beta_{\rm threshold}$  increases rapidly for higher $|\lambda/M^2|$ since larger $|\lambda/M^2|$ leads to faster and stronger development of the scalar field. Thus, one has to increase $\beta$ in order to saturate the scalar field at a lower value and keep the system in the hyperbolic region. One can also verify that for each $a_0/M$ the ratio $\lambda/(M^2\sqrt{\beta_{\rm threshold}})$ is roughly a constant as seen in the last column of Table \ref{tab:threshold}. This empirical trend can be understood by considering the coupling function used, for which the maximum value of $f'(\varphi)$ (for any $\varphi$) is limited from above for positive $\beta$ and scales as $f'(\varphi)_{\rm max}\sim 1/\sqrt{\beta}$. Thus the quantity $\lambda/(M^2\sqrt{\beta_{\rm threshold}})$ acts as an effective coupling at the threshold. We see that $\lambda/(M^2\sqrt{\beta_{\rm threshold}})$ decreases for larger $a_0/M$, which is expected since the absolute value of ${\cal R}^2_{GB}$ at the pole increases (recall that the negative ${\cal R}^2_{GB}$ around the poles of the apparent horizon is the source of the spin-induced scalarization).

	\begin{table*}
		\centering
		\caption{Threshold for loss of hyperbolicity calculated for several combinations of the initial angular momentum $a_0/M$ and $\lambda/M^2$. The third column represents the minimum $\beta_{\rm threshold}$ for which the simulations are still hyperbolic while the last one is the ratio $\lambda/(M^2\sqrt{\beta_{\rm threshold}})$ at the threshold, which is roughly a constant for every value of $a_0/M$.}
			\begin{tabular}{|c|c|c|c|}
			\hline
				$a_0/M$ & $\lambda/M^2$ & $\beta_{\rm threshold}$ & $\lambda/(M^2\sqrt{\beta_{\rm threshold}})$ \\
			\hline
				0.6 & -200 & 1000 & -6.3\\ 
				\hline
				0.6 & -400 & 5000 & -5.7\\
				\hline
				0.6 & -800 & 20950 & -5.5\\
				\hline
				\hline
				0.8 & -50 & 240 & -3.2\\
				\hline
				0.8 & -100 & 1150 & -3.0\\
				\hline
				0.8 & -200 & 4650 & -2.9\\
			\hline
					
			\end{tabular}
		\label{tab:threshold}
	\end{table*}

\begin{figure}[]
		\includegraphics[width=1\textwidth]{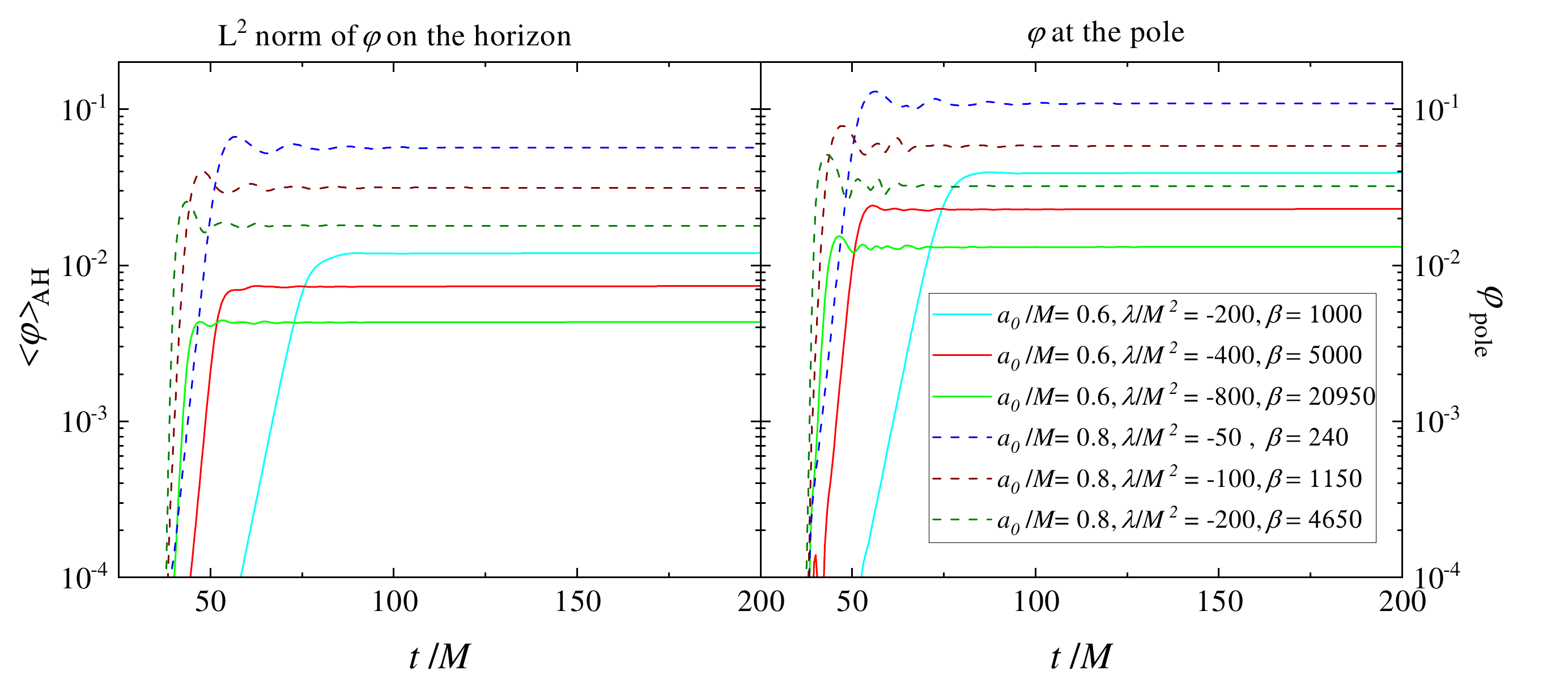}
		\includegraphics[width=1\textwidth]{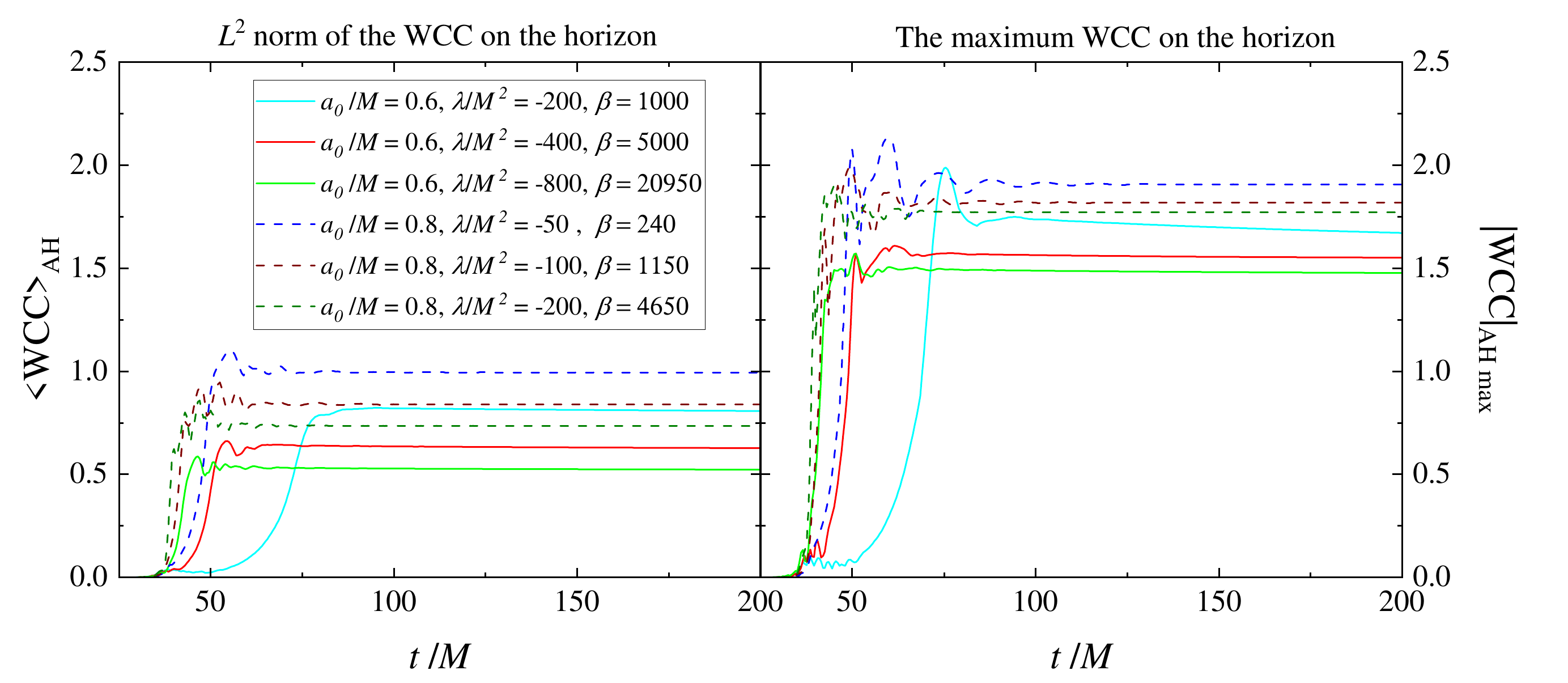}
		\caption{Time evolution of the models from Table \ref{tab:threshold} that are just above the hyperbolicity loss threshold. \textit{Top:} The scalar field at the apparent horizon taken either as the $L^2$ norm (\textit{left}) or at the pole of the apparent horizon, where it has its maximum (\textit{right}). \textit{Bottom:} The weak coupling condition at the apparent horizon taken  as the $L^2$ norm (\textit{left}) and its maximum value at the apparent horizon (\textit{right}).}
		\label{fig:Evol_threshold}
	\end{figure}

    For the threshold models listed in Table \ref{tab:threshold}, in Fig. \ref{fig:Evol_threshold} we show the evolution of the scalar field on the horizon as well as the weak coupling condition defined in Eq. \eqref{wfc}. Both quantities are displayed as their $L^2$ norm at the apparent horizon in the left panel. In the right panel, we depict the time evolution of the maximum value of the scalar field and the weak coupling condition. For the scalar field $\varphi$, this maximum typically occurs at the pole of the apparent horizon where the Gauss-Bonnet invariant is most negative. We, therefore, plot $\varphi$ at the pole. The extremum of the WCC happens away from the pole and its position slightly varies with time; we plot its maximum value on the apparent horizon as a function of time. The profiles of these quantities outside the black hole horizon will be shown in more detail in the next subsection.
         
    Let us first consider the scalar field evolution in the top panels of Fig. \ref{fig:Evol_threshold}. We see that the scalar field behavior (specifically its $L^2$ norm across the apparent horizon and at the pole of the apparent horizon) is very similar for all cases -- after a phase of exponential growth, small amplitude oscillations are observed until the scalar field settles to a constant profile. The value of the scalar field at the pole of the horizon $S^2$, which corresponds to the maximum, is roughly twice the size of the averaged  $L^2$ norm. 
        
    The differences between the plots in the left and the right panel are more pronounced for the weak coupling condition. In order for the EFT approach to be justified, the weak coupling condition defined in \eqref{wfc} should be much less than unity. As one can see in the left panel, ${\rm WCC}\lesssim 1$ for all threshold models if one looks at the $L^2$ norm of the WCC. The maximum value of the WCC can reach much larger values though, more than twice the average ones. In addition, at intermediate times, as the scalar field is still rapidly evolving, we observe spikes in the WCC, especially in its maximum value on horizon. 

	\begin{figure}[]
		\includegraphics[width=1\textwidth]{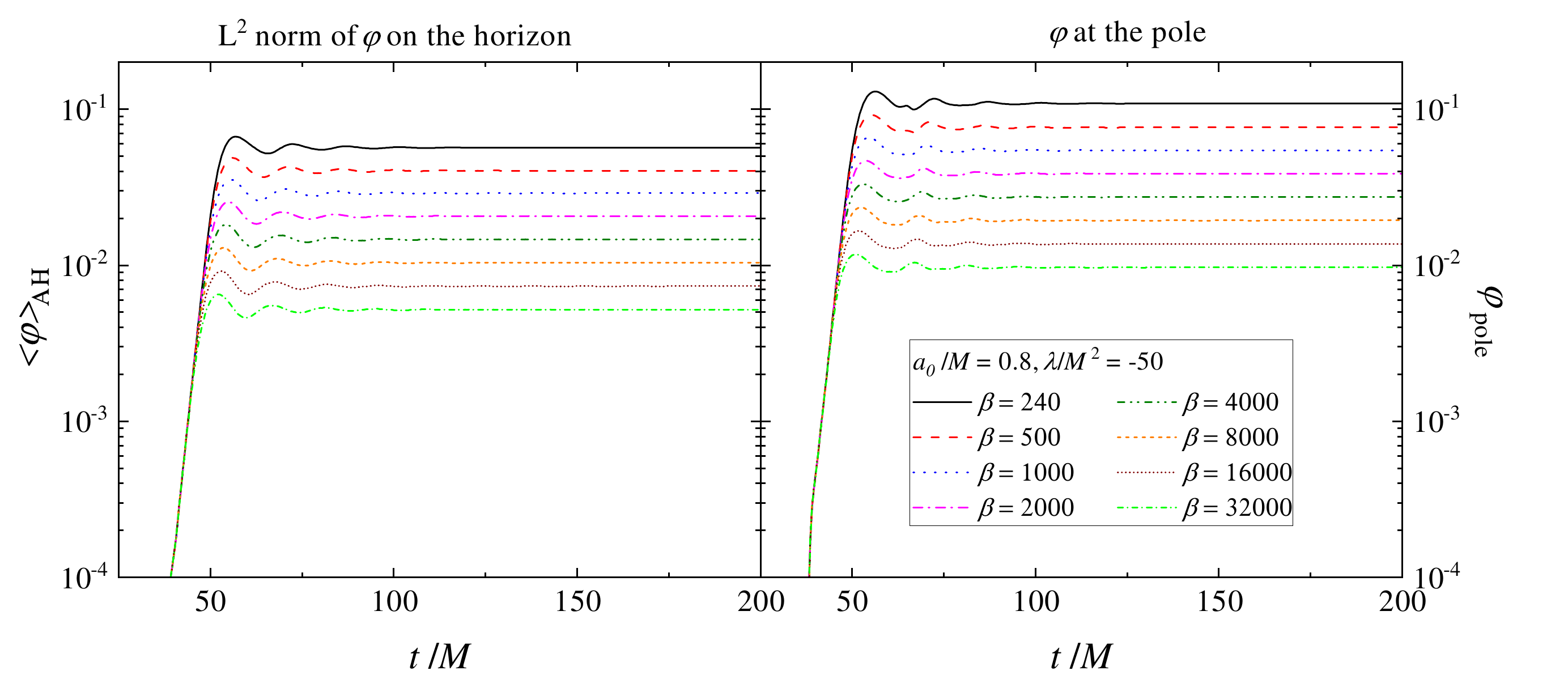}
		\includegraphics[width=1\textwidth]{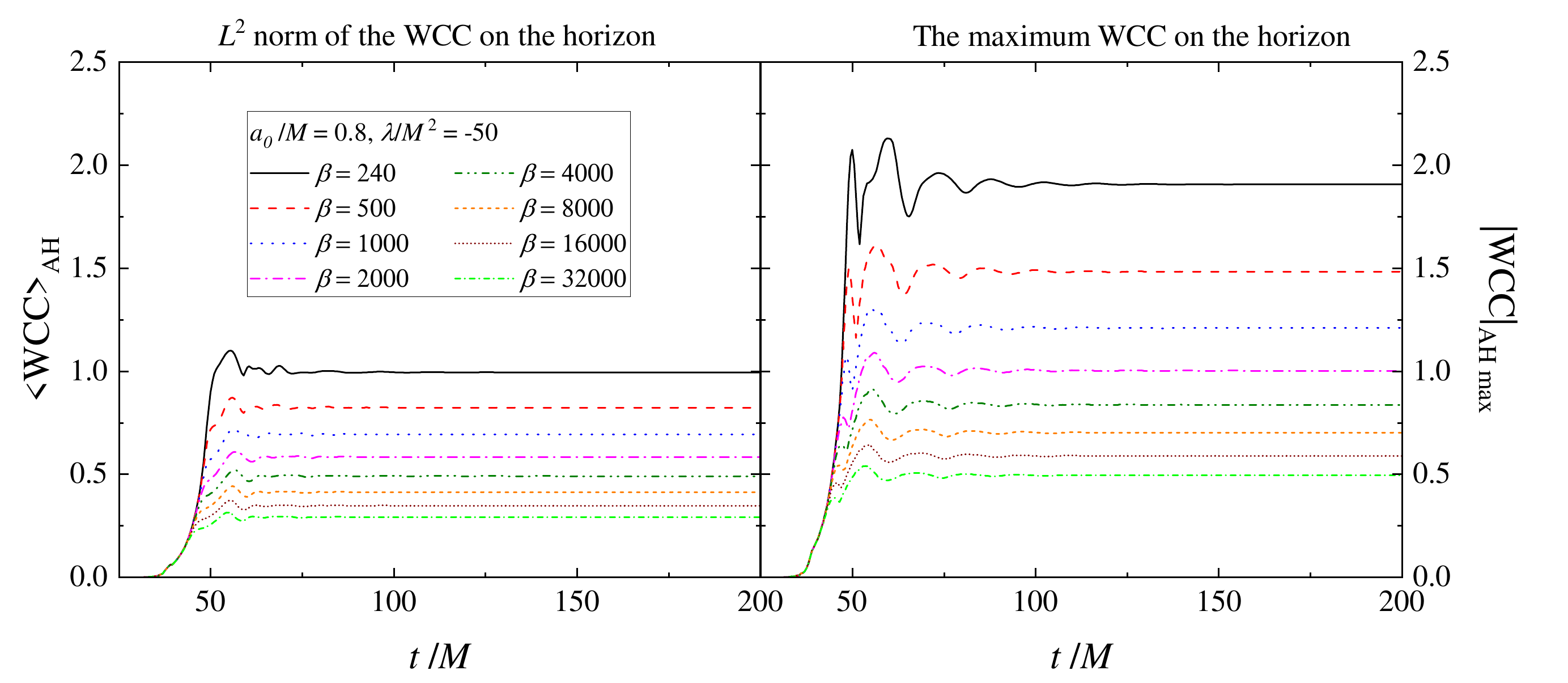}
		\caption{Time evolution of models with $a_0/M=0.8$, $\lambda/M^2=-50$ and varying $\beta$ starting from the threshold for hyperbolicity loss in Table \ref{tab:threshold} and reaching values for which the weak coupling condition is much less than unity. \textit{Top:} The scalar field at the apparent horizon taken either as the $L^2$ norm (\textit{left}) or at the pole, where it has a maximum (\textit{right}). \textit{Bottom:} The weak coupling condition at the apparent horizon taken  as the $L^2$ norm (\textit{left}) and its maximum value at the apparent horizon (\textit{right}).}
		\label{fig:Evol_Fixed_a_beta}
	\end{figure}

    We conclude that the range of parameters where hyperbolicity breaks down is actually well into the regime where the weak coupling condition is violated and the effective field theory treatment is no longer justified. We can explore how large $\beta$ should be for fixed $\lambda/M^2$ and $a_0/M$ in order to maintain the maximum WCC value at less than unity. For that purpose, we have chosen one combination of parameters, namely $a_0/M=0.8$ and $\lambda/M^2=-50$, and increased $\beta$ starting from its threshold value given in Table \ref{tab:threshold}. In order to maintain WCC$\ll 1$, $\beta$ needs to be at least two orders of magnitude larger than the minimum one that preserves hyperbolicity. The effect on the scalar field magnitude is less pronounced, and the two limiting cases in the figure, $\beta=240$ and $\beta=32000$, have roughly one order of magnitude difference in the equilibrium value of the scalar field after saturation. 
         
    \subsection{Illustration of diagnostics on 2D slices}

    In this subsection, we will present the evolution of the three main diagnostic quantities -- the scalar field, the WCC and the determinant of the effective metric, in the form of 2D slice plots. This provides further context on the hyperbolicity loss and WCC violation. In the figures below we will plot the normalized determinant 
    \begin{equation} \label{eq:Geff}
            G_{\rm eff} =  \frac{\det(g^{\mu\nu}_{\text{eff}})}{\det(g^{\mu\nu})} \left(1+\Omega^{\perp\perp}\right)^2 ~,
    \end{equation}
    which typically has values of order one and is normalized to unity in the absence of any scalar field.
        
\begin{figure}[]
	\includegraphics[width=1\textwidth]{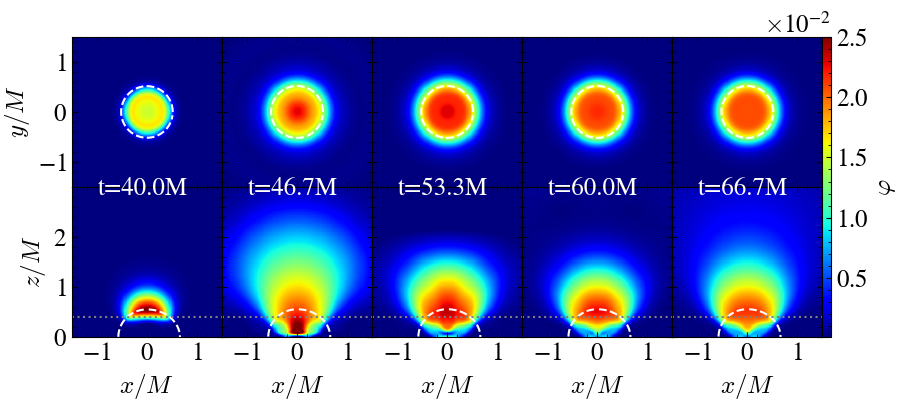}
	\caption{The scalar field profile at several equally spaced coordinate times during the scalarization. In the upper panels we have represented the scalar field taken on a $x-y$ slice, while in the lower panels $x-z$ slices are depicted. The $x-z$ slices cross the center of the BH while the $x-y$ slices are taken at $z=0.4$ above the center (marked with grey dotted lines in the lower panels), where the scalar field is strongest. The dashed white line represents the position of the apparent horizon. Note that in the black hole interior the Gauss-Bonnet term is completely turned off, thus the inner region does not represent a solution of the sGB field equations. The values of the employed parameters are $a_0/M=0.8$, $\lambda/M^2=-200$, and $\beta=10000$.}
	\label{fig:2D_Slice_phi}
    \end{figure}

\begin{figure}[]
	   \includegraphics[width=1\textwidth]{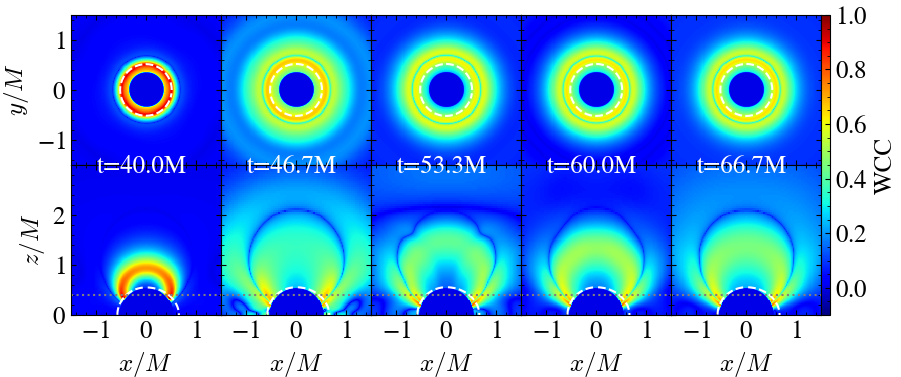}
	   \caption{The same configuration as in Fig. \ref{fig:2D_Slice_phi} but illustrating the evolution of the weak coupling condition.}
	   \label{fig:2D_Slice_WFC}
    \end{figure}

\begin{figure}[]
	   \includegraphics[width=1\textwidth]{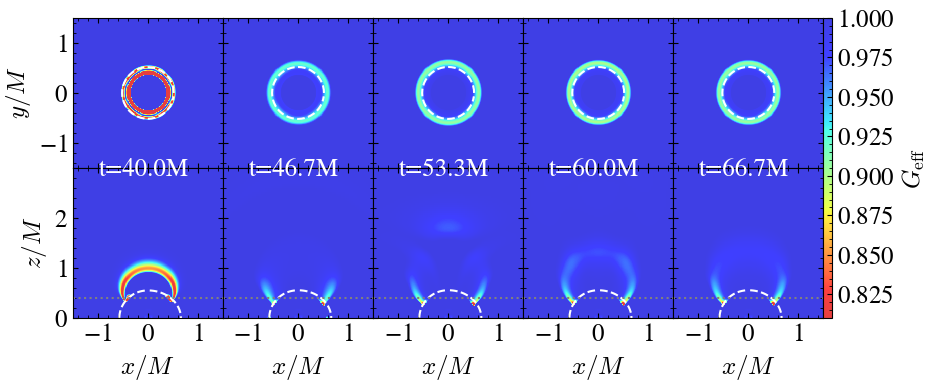}
	   \caption{The same configuration as in Fig. \ref{fig:2D_Slice_phi} but illustrating the evolution of the normalized determinant of the effective metric $G_{\rm eff}$, as defined in eq. \eqref{eq:Geff}.}
	   \label{fig:2D_Slice_discr}
    \end{figure}

    In Figs. \ref{fig:2D_Slice_phi}, \ref{fig:2D_Slice_WFC}, and \ref{fig:2D_Slice_discr} we have depicted $x-y$ and $x-z$ slices of the three quantities. Five equally spaced characteristic times during the evolution are represented starting from the moment when the initial scalar field seed pulse has arrived at the black hole (when the scalar hair starts growing), until it settles into an equilibrium profile. On each figure, the position of the apparent horizon is indicated with a dashed white line. The simulations shown correspond to the case of initial angular momentum $a_0/M=0.8$, $\lambda/M^2=-200$, and $\beta=10000$.

    The scalar field depicted in Fig. \ref{fig:2D_Slice_phi} has a maximum on the rotational axis and it forms a bulb shape there, while close to the equator it has a minimum. The WCC has a different distribution as seen in Fig. \ref{fig:2D_Slice_WFC} -- it has a maximum at a ring-like structure around the pole. This is because the maximum of the WCC is influenced not only by the scalar field maximum but also by its derivatives.

    Let us now consider the normalized determinant of the effective metric depicted in Fig. \ref{fig:2D_Slice_discr}. As discussed above, the moment when hyperbolicity is lost is associated with the determinant becoming zero.
    Since inside the black hole interior we turn off the Gauss-Bonnet coupling term, the value of the determinant is unity there. The values of $a_0/M$, $\lambda/M^2$, and $\beta$ are chosen in such a way that the models we evolve are above the threshold for loss of hyperbolicity, so the determinant remains positive outside the apparent horizon.  It reaches a minimum value (closer to hyperbolicity loss) at intermediate times, when the scalar field grows fastest. At that point of evolution, the WCC is also at its maximum. At late times the determinant approaches larger values (close to unity), that is, further away from hyperbolicity loss. This effect is expected since both the determinant and the WCC are influenced by the scalar field time and spatial derivatives. Thus, the loss of hyperbolicity will depend not only on the final stationary solution but also on the dynamical evolution that leads to it.

    In our simulations of models with smaller $\beta$, in a regime for which the evolution is stable but closer to the threshold for loss of hyperbolicity, we observe a transient period where hyperbolicity is lost within the horizon. Specifically, inside the apparent horizon, but still outside the region where the Gauss-Bonnet coupling is turned off, the determinant has negative values at intermediate times. Since this happens inside the apparent horizon for a relatively short time, the evolution manages to continue. This is consistent with the observations made in the $1+1$ nonlinear evolution of \cite{Corelli:2022phw}.

    \subsection{Gauge dependence of hyperbolicity loss}\label{subsec:gauge}
    
    As discussed in the previous sections, the breakdown in hyperbolicity that we observe appears to be linked to the physical modes.  These are gauge invariant \cite{Reall:2021voz}, and so we do not expect a change in gauge to prevent the breakdown observed. However, it is possible that in the strongly coupled regime the gauge modes themselves may become problematic. It is also interesting to explore the effect of the modified puncture gauge choice (which is newly developed and little explored) on the accuracy and resolution of the numerical simulations. For that purpose, in this section we explore the impact of modifying the gauge, by adjusting the free functions $a(x)$ and $b(x)$ that determine the auxiliary metrics (see eq. \eqref{eq:guage_choice}). 
        
    In previous sections we have worked with constant $a(x)=0.2$ and $b(x)=0.4$, values similar to those used in \cite{East:2021bqk}. This choice is by no means unique, and we have tried several other combinations of constant $a(x)$ and $b(x)$, keeping always $b(x)>a(x)$ and $\kappa_2>-\tfrac{2}{2+b(x)}$ as required in the modified puncture gauge. We focus on simulations of black holes with $a_0/M=0.8$ and $\lambda/M^2=-50$. The combinations of $\{a(x),b(x)\}$ we have tested are $\{0.05,0.1\}$, $\{0.1,0.2\}$, $\{0.2,0.4\}$, $\dots$, $\{1.6,3.2\}$, as well as $\{0.2,0.3\}$ and $\{0.2,0.6\}$. For all cases, the threshold for loss of hyperbolicity is found consistently to be at $\beta=240$, with any deviations lying within our numerical uncertainties. A practical observation is that, despite the wide range of $\{a(x),b(x)\}$ that we have considered, we were able to perform evolutions for all of the combinations mentioned above, keeping the same value of the damping and dissipation parameters, $\kappa_1,\;\kappa_2$ and $\sigma$ respectively,  as for the $\{0.2,0.4\}$ case.
             
    \begin{figure}[]
	   \includegraphics[width=0.5\textwidth]{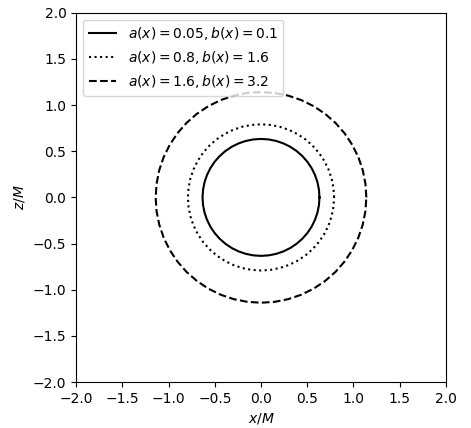}
	   \caption{An $x-z$ slice of the apparent horizon at $t=400M$ once the BH has settled to a stationary gauge. The simulations are performed for  $a_0/M=0.8$, $\lambda/M^2=-50$, $\beta=2000$ and different combinations of the gauge parameters $a(x)$ and $b(x)$. }
		\label{fig:horizon_gauge}
	\end{figure}
 
    We now highlight another interesting observation not directly related to the loss of hyperbolicity. While experimenting with the values of $a(x)$ and $b(x)$ we have noticed that changing their values in turn changes significantly the coordinate size of the apparent horizon radius. For example the difference between the two limiting cases we have considered, namely $\{a(x),b(x)\}=\{0.05,0.1\}$ and $\{a(x),b(x)\}=\{1.6,3.2\}$ is roughly twice, as evident from Fig. \ref{fig:horizon_gauge}. The coordinate radius of the apparent horizon increases further with the increase of $a(x)$ and $b(x)$. This might be helpful for some simulations, as it may provide a way to improve the resolution at the apparent horizon. Another positive outcome concerns the violation of the Hamiltonian constraint, that always occurs near the puncture in singularity avoiding coordinates. When we increase $a(x)$ and $b(x)$ the size of this constraint violating region grows slower compared to the growth of the apparent horizon radius. As a result, in the simulation with  $\{a(x),b(x)\}=\{1.6,3.2\}$, the constraint violating region is further away from the black hole apparent horizon compared to the standard $\{a(x),b(x)\}=\{0.2,0.4\}$ case. This should improve the accuracy of the evolution outside the horizon for a fixed resolution, since the separation (in terms of grid points) between the apparent horizon and the region with sizeable constraint violations is increased. 

    These results, whilst not definitive, suggest that the hyperbolicity loss that we observe does not strongly depend on the choice of the functions $a(x)$ and $b(x)$, and is dominated by the physical sector of the eigenmodes (rather than the gauge modes). This also follows from our observation that the breakdown of the simulations is usually preceded by the determinant of the effective metric turning negative, which indicates that some of the eigenvalues from the physical sector have become complex. Thus, the results are consistent with the theoretical investigations in \cite{Reall:2021voz}.

    \subsection{Reasons for loss of hyperbolicity}
    
    \begin{figure}[]
		\includegraphics[width=1\textwidth]{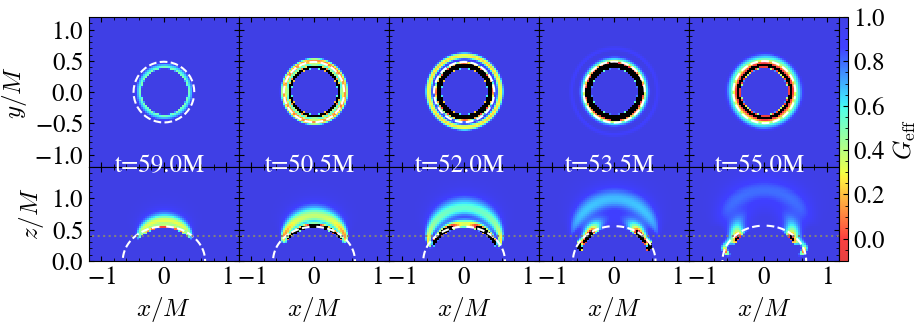}
		\includegraphics[width=1\textwidth]{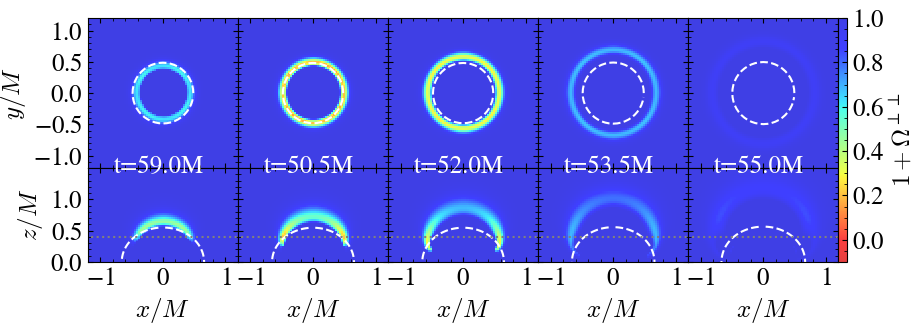}
		\caption{Time evolution of a model just below the threshold of hyperbolicity loss with $a_0/M=0.8$, $\lambda/M=-50$, and $\beta=200$. Several equally spaced coordinate times during the scalarization are plotted, capturing the evolution just before the code breaks down due to hyperbolicity loss. In each figure, both $x-y$ and $x-z$ slices are depicted, as in the figures above. The apparent horizon is plotted as a white dashed line. \textit{(top)} Time evolution of the normalized determinant of the effective metric $G_{\rm eff}$ defined by eq. \eqref{eq:Geff}. Negative values are depicted in black. \textit{(bottom)} Time evolution of $1+\Omega^{\perp\perp}$. Negative values are depicted in black.}
		\label{fig:discriminant_beta200}
	\end{figure} 

    \begin{figure}[]
		\includegraphics[width=1\textwidth]{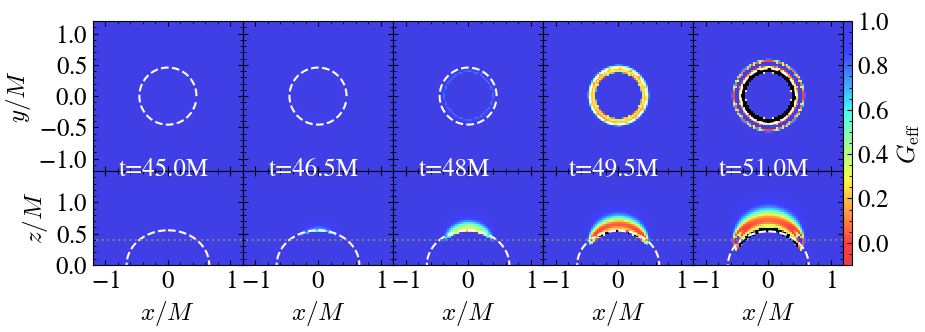}
		\includegraphics[width=1\textwidth]{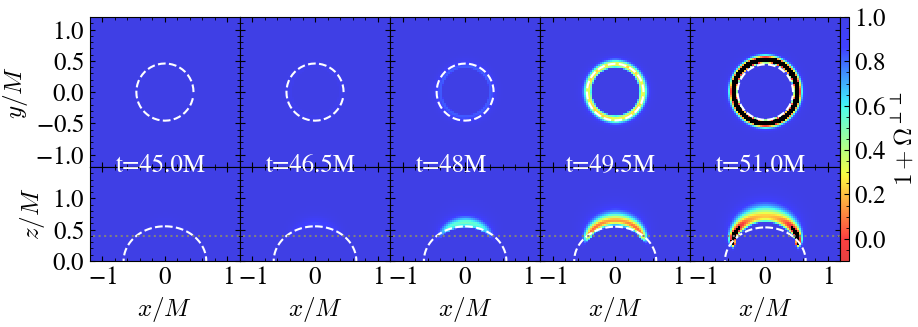}
		\caption{As in Fig. \ref{fig:discriminant_beta200} but for a model far beyond the boundary of hyperbolicity loss with $a_0/M=0.8$, $\lambda/M=-50$, and $\beta=100$. Negative values of the determinant of the effective metric and the quantity  $1+\Omega^{\perp\perp}$ are depicted in black.}
		\label{fig:discriminant_beta100}
	\end{figure}
 
    In this subsection we investigate whether the hyperbolicity loss happens because of a Tricomi or a Keldysh-type of transition. Since the eigenvalues that we can directly compute lie at the null cone of the effective metric, one can show that their propagation speeds will only diverge if the quantity $1+\Omega^{\perp\perp}$ vanishes, thus this indicates a Keldysh-type hyperbolicity loss. If instead the determinant of the effective metric vanishes with the quantity $1+\Omega^{\perp\perp}$ remaining positive, the propagation speeds remain finite but they become degenerate, which indicates a Tricomi-type hyperbolicity loss.
        
    In order to test the type of transition, we have performed two simulations that lose hyperbolicity at a certain point with fixed $a_0/M=0.8$, $\lambda/M^2=-50$, and varying $\beta=\{100,200\}$. While  $\beta=200$ is only slightly below the threshold given in Table \ref{tab:threshold}, for $\beta=100$ the hyperbolicity loss happens faster with the formation of a larger region of a negative determinant of the effective metric. 
        
    The quantity $1+\Omega^{\perp\perp}$  behaves differently in the two cases. For $\beta=200$, it is always positive as seen in Fig. \ref{fig:discriminant_beta100}. Thus, the hyperbolicity loss is caused by a Tricomi-type transition, as was observed in other simulations in sGB gravity \cite{Ripley:2019hxt}.  Note that the normalized determinant of the effective metric, depicted in the same figure, is negative only in a small region outside the apparent horizon. As may be expected, shortly after this region forms, the numerical evolution cannot be continued.  
        
    As depicted in Fig. \ref{fig:discriminant_beta100}, for $\beta=100$ the quantity $1+\Omega^{\perp\perp}$ can become negative outside the apparent horizon, which means that the propagation speeds diverge. Interestingly, in the region where this happens, the determinant is positive. Thus, one can speculate that in the region with a negative determinant (in black in the upper panel of Fig. \ref{fig:discriminant_beta100}, around the pole of the apparent horizon), the hyperbolicity loss is again of the Tricomi-type since the propagation speeds are finite there. However, the appearance of another region with diverging speeds (in black in the bottom panel of Fig. \ref{fig:discriminant_beta100}) implies that a second ``problematic'' region develops away from the pole, causing a loss of hyperbolicity of the Keldysh-type. Further investigation is required to confirm that the loss of hyperbolicity happens because Tricomi-type and Keldysh-type-of transitions occur simultaneously in different regions of the spacetime, or whether one is dominant.

    \section{Conclusions}
    
    In this work we performed $3+1$ nonlinear evolutions of rotating black holes in sGB theories of gravity that admit spin-induced scalarization. This is a mechanism to endow black holes with scalar hair, with the trigger for the scalar field development being a sufficiently fast black hole rotation ($a/M>0.5$) and a sufficiently strong coupling $\lambda$. The $3+1$ evolution is performed with a modified version of the public \texttt{GRChombo} code and the newly developed modified CCZ4 gauge. Apart from the scalar field development, we monitored two important quantities. The first is the determinant of the effective metric, which, if smaller than zero, indicates that the hyperbolicity of the field equations is lost. The second one is the weak coupling condition, which, if larger than one, indicates that treating sGB gravity as an effective field theory is no longer justified. We investigated in detail the threshold and dynamics of hyperbolicity loss, the influence of our chosen gauge, and how these issues connect to the regime of validity of the model as an effective field theory.

    We focused on two initial black hole angular momenta -- $a_0/M=0.6$, which is close to the threshold for the development of spin-induced scalarization, and $a_0/M=0.8$, which is a sufficiently fast rotation that rapid scalar field development occurs even for moderate values of the coupling parameters. The initial data is a Kerr black hole with a small scalar field pulse superimposed on it. For a particular coupling function, being quadratic in the scalar field at leading order, we have determined the threshold for hyperbolicity loss. We find that the determinant of the effective metric reaches its smallest values in the early period of scalar field growth, before an equilibrium state is reached. This means that hyperbolicity is typically lost during the formation of the scalar cloud.
    We have also examined whether the loss of hyperbolicity happens because of a Tricomi or a Keldysh-type of transition. For models close to the scalarization threshold, it is clearly of Tricomi-type (i.e., some characteristic speeds go to zero). On the other hand, if we consider models deep inside the strongly coupled regime, i.e., having faster and stronger scalar field development, the transition appears to change to a Keldysh-type (i.e., some characteristic speeds diverge) or a mixture of both developing in different regions of the spacetime.

    The physical modes are gauge invariant \cite{Reall:2021voz} and so the breakdown we observe should persist for other gauge choices. However, it is possible that a change of gauge or formulation may create additional instabilities in the strong coupling regime. We have not, however, seen any change in behaviour for the gauge choices we have considered. Methods such as ``fixing'' should make a difference to the physical modes because they change the theory, introducing new degrees of freedom, see \cite{Franchini:2022ukz}, but we do not investigate such methods here.
    Performing an exhaustive analysis of all gauges is not possible, but we have checked that simple gauge changes do not influence the hyperbolicity loss threshold, which is consistent with \cite{Reall:2021voz}. Interestingly, the change of gauge can have other benefits, even in pure GR, such as an increase of horizon radius that can potentially lead to better resolution at the apparent horizon. The constraint violation region inside the black hole also shrinks relative to the black hole size for those gauge choices. 
        
    If we consider sGB gravity not as a standalone theory, but instead as an EFT that arises as a low energy limit of a more fundamental theory, the physically relevant question is at what point during the evolution the theory exits the domain of validity of EFT, i.e, when higher-derivative corrections cannot be neglected. In order to quantify this we defined a weak coupling condition (WCC). Our results show that this condition reaches values much larger than unity for models on the threshold of scalarization (these are typically the models with the strongest deviation from GR). Keeping the WCC much less than one in the simulations requires adjustment of the coupling parameters so that the resulting scalar field is roughly an order of magnitude smaller than the scalar field of the hyperbolicity loss threshold model.
        
    These results are consistent with the results in the alternative mGHC gauge \cite{East:2021bqk}, and indicate that spin-induced scalarization leads to a similar violation of hyperbolicity as the case of standard scalarization. The effect appears general, and only weakly dependent on the particular flavor of sGB gravity and the source of the scalar field. However, the more physical condition is the WCC, which we have consistently found to be strongly violated prior to loss of hyperbolicity for the class of initial conditions that we have considered. Therefore, as long as one considers sGB to be an EFT and only applies it within its regime of validity, a consistent and stable evolution should be achievable.

    \section*{Acknowledgements}
    This study is in part financed by the European Union-NextGenerationEU, through the National Recovery and Resilience Plan of the Republic of Bulgaria, project No. BG-RRP-2.004-0008-C01. DD acknowledges financial support via an Emmy Noether Research Group funded by the German Research Foundation (DFG) under grant no. DO 1771/1-1. We acknowledge Discoverer PetaSC and EuroHPC JU for awarding this project access to Discoverer supercomputer resources. We thank the entire \texttt{GRChombo} \footnote{\texttt{www.grchombo.org}} collaboration for their support and code development work. PF and KC are supported by an STFC Research Grant ST/X000931/1 (Astronomy at Queen Mary 2023-2026). PF is supported by a Royal Society University Research Fellowship  No. URF\textbackslash R\textbackslash 201026, and No. RF\textbackslash ERE\textbackslash 210291. KC is supported by an STFC Ernest Rutherford fellowship, project reference ST/V003240/1. LAS is supported by a QMUL Ph.D. scholarship. 
    Development of the code used in this work utilised the ARCHER2 UK National Supercomputing Service\footnote{\texttt{https://www.archer2.ac.uk}} under the EPSRC HPC project no. E775, the CSD3 cluster in Cambridge under Projects No. DP128. The Cambridge Service for Data Driven Discovery (CSD3), partially operated by the University of Cambridge Research Computing on behalf of the STFC DiRAC HPC Facility. 
    The DiRAC component of CSD3 is funded by BEIS capital via STFC capital Grants No. ST/P002307/1 and No. ST/ R002452/1 and STFC operations Grant No. ST/R00689X/1. DiRAC is part of the National e-Infrastructure\footnote{\texttt{www.dirac.ac.uk}}. 
    Calculations were also performed using the Sulis Tier 2 HPC platform hosted by the Scientific Computing Research Technology Platform at the University of Warwick. Sulis is funded by EPSRC Grant EP/T022108/1 and the HPC Midlands+ consortium. This research has also utilised Queen Mary's Apocrita HPC facility, supported by QMUL Research-IT.

    \appendix*
    \section{Code testing and convergence}
    
    The sGB modification of the \texttt{GRChombo} code was tested in \cite{AresteSalo:2023mmd}. In addition, for this work we have verified that the code produces the correct scalar field evolution in the case of isolated black holes both in the case of shift-symmetric sGB gravity and for sGB theories admitting spontaneous scalarization. Both the growth time of the scalar field and its equilibrium value match well to known results in \cite{Doneva:2017bvd,Blazquez-Salcedo:2018jnn,Doneva:2021dqn}. As a second step, the code was compared to results in the decoupling limit approximation for isolated rotating black holes \cite{Doneva:2021dqn}. A very good agreement was demonstrated both for the standard scalarization and the spin-induced one up to the level of accuracy in \cite{Doneva:2021dqn}.

\begin{figure}[]
	\includegraphics[width=0.6\textwidth]{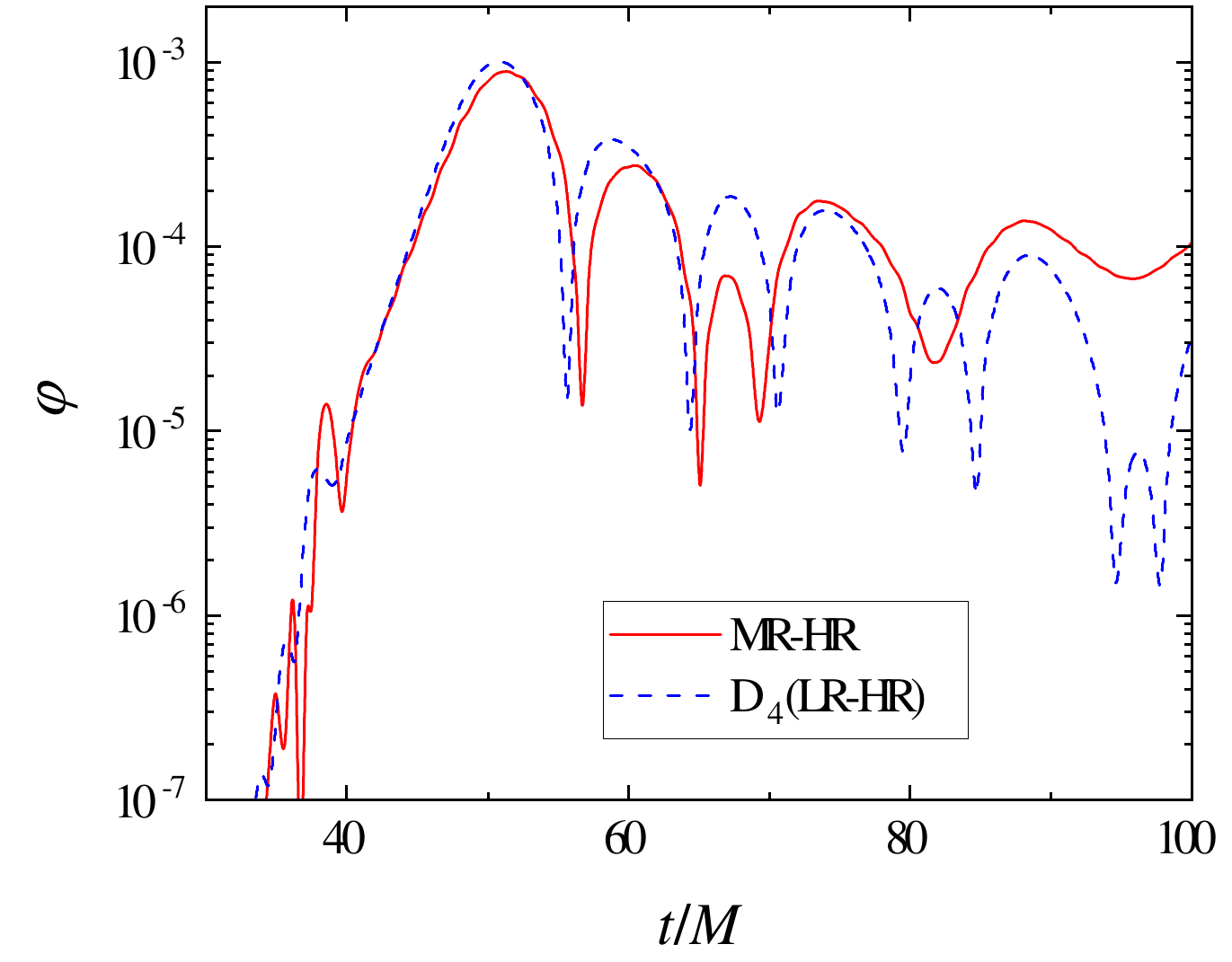}
	\caption{The difference between the scalar field time evolution performed for three different resolutions for $a_0/M=0.8$, $\lambda/M^2=-50$, and $\beta=500$. The solid red line shows the difference between medium and high resolution while the blue dashed line is the difference between low and medium resolution multiplied by the fourth-order convergence factor $D_4=\frac{h^4_{\rm LR} - h^4_{\rm ML}}{h^4_{\rm MR} - h^4_{\rm HR}}$.}
	\label{fig:Convergence}
\end{figure}

    Apart from testing against known results, we have also performed convergence tests. Namely, we have followed the black hole evolution for fixed $a_0/M=0.8$, $\lambda/M^2=-50$, $\beta=500$, and three different resolutions. A comparison between the results for the scalar field development in the three cases is shown in Fig. \ref{fig:Convergence}. As seen, the convergence matches well to fourth order as also expected from \cite{AresteSalo:2022hua}. 

\bibliography{references}

\end{document}